\documentclass[a4paper]{article}

\usepackage{amsmath,amssymb,subeqnarray}
\usepackage{graphicx}
\usepackage{natbib}

\newcommand{\dd}{\mathrm{d}}
\newcommand{\Nl}{\,\mathrm{N}}
\newcommand{\Wl}{\,\mathrm{W}}

\newcommand{\mean}[1]{\left<#1\right>}

\newcommand{\ord}[1]{O(#1)}

\newcommand{\fT}{\phi_\theta^{T}}
\newcommand{\fW}{\phi_\theta^{d}}
\newcommand{\fWc}{\phi_\theta^{*}}
\newcommand{\zs}{z^*}
\newcommand{\Pc}{\mathcal{P}}
\newcommand{\s}{\nobreak\mbox{$\;$s}}

\newcommand{\m}{\nobreak\mbox{$\;$m}}
\newcommand{\kg}{\nobreak\mbox{$\;$kg}}

\newcommand{\dC}{\nobreak\mbox{$\;$K}}
\newcommand{\Hz}{\nobreak\mbox{$\;$Hz}}

\begin{document}

\title{Estimates of the temperature flux--temperature gradient relation above a sea-floor}

\author{Andrea A. Cimatoribus\thanks{Email address for correspondence: \mbox{Andrea.Cimatoribus@nioz.nl}}, H. van Haren\\
\small Royal Netherlands Institute for Sea Research, \\
\small Landsdiep 4, 1797SZ, 't Horntje, NH, the Netherlands}
\date{\today}

\maketitle

\begin{abstract}
The relation between the flux of temperature (or buoyancy), the vertical temperature gradient and the height above the bottom, is investigated in an oceanographic context, using high-resolution temperature measurements.
The model for the evolution of a stratified layer by \citet{balmforth_dynamics_1998} is reviewed and adapted to the case of a turbulent flow above a wall.
Model predictions are compared to the average observational estimates of the flux, exploiting a flux estimation method proposed by \citet{winters_diascalar_1996}.
This estimation method enables the disentanglement of the dependence of the average flux on the height above the bottom and on the background temperature gradient.
The classical N-shaped flux-gradient relation is found in the observations.
Model and observations show similar qualitative behaviour, despite the strong simplifications used in the model.
The results shed light on the modulation of the temperature flux by the presence of the boundary, and support the idea of a turbulent flux following a mixing-length argument in a stratified flow.
Furthermore, the results support the use of Thorpe scales close to a boundary, if sufficient averaging is performed, suggesting that the Thorpe scales are affected by the boundary in a similar way as the mixing length.
\end{abstract}

\section{Introduction}
\label{sec:introduction}

Flows in geophysical fluids span a remarkably broad range of scales.
From a practical point of view, the interest concentrates on the large scales, $\ord{10^3-10^5\m}$, most relevant for our experience of e.g.~the climate, the concentration of pollutants, the ocean productivity.
Geophysical flows are, however, highly non-linear, implying that the behaviour of the large scales is not independent of the small ones, $\ord{10^{-3}-10^2\m}$.
This is one of the main motivations underlying the study of small-scale turbulence in geophysical flows.

In this work, the vertical turbulent transport of a scalar, namely temperature, is investigated, using observations collected above a sloping sea-floor.
The specific interest for benthic boundary layers, sloping in particular, is connected to the suggestion \citep{munk_abyssal_1966} that the regions directly above the sea-floor dominate the globally averaged vertical buoyancy transport in the ocean.
Here, we address this issue from the point of view of the relation between the flux of temperature ($\phi_\theta$), the gradient of temperature, and the distance above the solid boundary.
The temperature flux is related to the heat flux by a factor $\rho c$, with $\rho$ the density and $c$ the specific heat.
By considering the temperature instead of the heat flux, we assume that density and specific heat are approximately constant.

The flux--gradient relation has received attention in particular because it may explain the intermittency of density gradients in stratified turbulence, i.~e.~the presence of layers.
Layering is ubiquitous in the observations we present here, as in most observations from geophysical flows.
\citet{phillips_turbulence_1972} and \citet{posmentier_generation_1977} hypothesised that the time evolution of buoyancy in a stratified flow could be modelled, assuming horizontal homogeneity, by a diffusion equation of the form:
\begin{equation}
  b_t= \left(\phi(b_z)\right)_z,
\end{equation}
where the subscripts indicate derivation, $b$ is the buoyancy, $\phi(b_z)$ is the buoyancy flux, seen as a function of the buoyancy gradient, $t$ is time and $z$ is the vertical coordinate.
If the flux function $\phi(b_z)$ is non-monotonic, an instability can occur for $\dd\phi(b_z)/\dd{b_z}<0$, leading to the unbounded growth of sharply stratified sheets divided by weakly stratified layers.
This mechanism for layer formation was later recognised in the laboratory by \citet{ruddick_formation_1989} and \citet{park_turbulent_1994}.

Building upon this idea, \citet{balmforth_dynamics_1998} (hereafter B98) proposed a model for the evolution of a stratified turbulent flow, again assuming horizontal homogeneity, in terms of two partial differential equations for the turbulent kinetic energy density ($e$) and the buoyancy gradient ($g=b_z$).
Despite its simplicity, the model is remarkably successful in describing the formation and evolution of layers in laboratory experiments.
The fundamental advance in B98 was the recognition that $\dd{\phi(g)}/\dd{g}$ must be greater than zero for large $g$, in order to stop the growth of the sharply stratified layers before they grow infinitely thin and strong, i.e.~the $\phi(g)$ function must be ``N-shaped'' (see e.g.~figure \ref{fig:model}, black lines).
B98 argued that this implies a forcing term in the kinetic energy equation proportional to a velocity scale (squared) divided by the eddy turnover time scale, leading to the correct N-shape of the equilibrium $\phi(g)$.

\citet{holford_turbulent_1999} further tested the model predictions in the laboratory, suggesting that the $N$-shape of $\phi(g)$ was the consequence rather than the cause of layer formation.
\cite{martin_layering_2006} also addressed the problem of layer formation in the laboratory, observing the development of an $N$-shaped $\phi(g)$ after layers formed.
\citet{wunsch_model_2001} reformulated the model in stochastic terms, also suggesting that the dynamics predicted by the model could be related to the zigzag instability \citep{billant_experimental_2000}.

Another important hypothesis in B98 is that $e$ and $b$ are transported by the turbulent flow according to a mixing-length argument, with the flux proportional to the gradient and to $l\,U$, where $l$ is a length scale and $U$ is a velocity scale.
In B98, $l$ was assumed to be the geometric average between a length scale proportional to the Ozmidov scale and a constant length scale $d$, set by the stirring device:
\begin{equation}
  l = \frac{d e^{1/2}}{\left(e+\gamma d^2 g\right)^{1/2}},
  \label{eq:mixing-length-b98}
\end{equation}
with $\gamma$ a non-dimensional constant.
The B98 model implies that the mixing length, relevant for computing the turbulent flux or turbulent diffusivity, is set by the Ozmidov scale under strong stratification, while it is set by $d$ under weak stratification.
The constant $\gamma$ controls the range where $l$ is influenced by the stirring device.

In this work, we will reinterpret the B98 model to describe the dependence of the buoyancy flux on the height above the bottom (HAB) and on the buoyancy gradient.
Assuming a mixing length proportional to the distance from the boundary is the approach in the classical problem of the ``law-of-the-wall'' for a constant-density fluid, see e.g.~chapter 5 of \cite{landau_fluid_1987} or the original discussion in \cite{prandtl_mechanics_1935}, section 21.
As in B98, we assume horizontal homogeneity but include the effect of the solid sea-floor, in a highly idealised way which also ignores the slope of the boundary (boundary normal and vertical directions thus coincide).
In doing so, we reinterpret $d$ in equation \ref{eq:mixing-length-b98} as a different length scale, namely the HAB $z$, which is the external length scale limiting the mixing length close enough to the solid boundary.
Differently from $d$, $z$ is not a constant.

The predictions of the model will be compared to estimates of the flux obtained from high resolution temperature observations above a sloping sea-floor.
The estimates are derived using the method suggested in \citet{winters_diascalar_1996}.
This method can be seen as an extension of the more common one by \citet{osborn_oceanic_1972}, and is based on the estimation of the small-scale buoyancy gradient, and on the adiabatic reordering of buoyancy profiles to remove inversions.
The method proposed in \citet{winters_diascalar_1996} has three main advantages over the one by \citet{osborn_oceanic_1972}: (i) it provides a clear definition of the background stratification to be used for estimating the flux, (ii) it provides an estimate of the flux due to molecular mixing (irreversible) rather than due to advection (reversible), and (iii) it provides a natural framework for studying local variations of the flux as a function of the background buoyancy gradient and HAB.

The flux estimates will also be compared to those obtained using the classical method of reordering buoyancy overturns, first outlined by \citet{thorpe_turbulence_1977}.
This method has been used in several observational studies \citep[see e.g.][]{dillon_vertical_1982,itsweire_measurements_1984,gregg_diapycnal_1987,gargett_ocean_1989}.
Despite recent criticism \citep{mater_quest_2014,chalamalla_mixing_2015}, it still can provide a reliable order-of-magnitude estimate of turbulence parameters, after sufficient averaging \citep{mater_biases_2015,chalamalla_mixing_2015}.
We also stress that the simulations of \citet{chalamalla_mixing_2015} are not directly relevant for the observations presented here, where a single overturn evolving from a laminar flow is never observed.
In the present work, we will show that the flux estimates based on overturn reordering are correlated with those from the method of \citet{winters_diascalar_1996}, despite a systematic low bias of the latter due to instrumental resolution limitations.
Furthermore, we will present a mixing-length argument supporting the use of the estimates from the reordering of overturns in the vicinity of a solid boundary, and provide observational evidence in favour of this argument.

The article is organised as follows.
In section \ref{sec:theory}, the extension of the model by B98 is briefly presented.
In section \ref{sec:dataset} the temperature observations are briefly described.
In sections \ref{sec:thorpe} and \ref{sec:winters-d'asaro}, the two methods used for estimating the flux are presented in some detail.
In section \ref{sec:results}, the flux estimates are presented and compared to the model.
Discussion of the results and conclusions follow in section \ref{sec:conclusions}.

\section{Theory}
\label{sec:theory}

We develop an extension of the B98 model with the aim of applying it to stratified turbulence above a sea-floor.
While the most notable feature of the B98 model is layer development, our interest in it is also motivated by its focus on the flux-gradient relation, which we can estimate in the observations.
Following the example of B98, the aim is to develop a model which captures the essential phenomenology in the simplest mathematical form.
The starting point are equations 2.1a, b of B98, which we rewrite here using a slightly modified notation:
\begin{subeqnarray}
  g_t = \left(l e^{1/2} g\right)_{zz},\\
  e_t = \beta\left(l e^{1/2} e_z\right)_z-l e^{1/2}g-\alpha l^{-1} e^{3/2}+\Pc.
  \label{eq:B98}
\end{subeqnarray}
We refer to B98 for a discussion and motivation of equations \ref{eq:B98}.
The vertical coordinate $z$ is zero at the bottom and increases upwards, $\alpha$ and $\beta$ are dimensionless constants, $\Pc$ is the energy production term.
Equations \ref{eq:B98} describe the coupled evolution of the buoyancy gradient and of the kinetic energy in a stratified turbulent flow, realistically enough to reproduce all the essential features of layer formation and evolution in the laboratory experiments discussed in B98.
Horizontal homogeneity is assumed, the slope of the sea-floor is thus ignored.
The possible impact of this far-fetched assumption is discussed in section \ref{sec:conclusions}.
Equations \ref{eq:B98} are considered here only for the highly idealised case of the steady state ($\partial/\partial t\equiv 0$).

We depart from B98 in the definitions of the mixing length $l$ and energy production term $\Pc$.
Along the same line of B98, the mixing length is defined as:
\begin{equation}
  l = \delta \frac{z e^{1/2}}{\left(e+\gamma z^2 g\right)^{1/2}},
  \label{eq:mixing_length}
\end{equation}
with $\delta$ a non-dimensional constant.
While $l$ has the same form as equation \ref{eq:mixing-length-b98}, the variable $z$ is used instead of the constant $d$.
As suggested already by \citet{thorpe_turbulence_1977} for the case of the surface boundary layer, at small $z$ the density overturns are limited by the presence of the boundary.
Here, we argue that the mixing length is similarly limited by the presence of the sea-floor, in analogy to the case of a wall-bounded boundary layer in a constant-density fluid \citep{landau_fluid_1987}.
The geometric average in \ref{eq:mixing_length} is chosen for reasons of mathematical simplicity.
We do not imply that this is the correct mixing length, but that \ref{eq:mixing_length} captures the essential characteristics of the system, namely that from a far field value $l\propto (e/g)^{1/2}$, the mixing length interpolates to $l\propto z$ close to the boundary.
Identifying the exact dependence of $l$ on $z$ and $g$ is beyond the scope of this work.
Since the observations suggest that the influence of the sea-floor extends throughout the mooring ($100\m$), the constant $\gamma$ is expected to be much smaller than 1, implying that the asymptotic value of $l$ is approached only above the top of the mooring.
For any realistic value of $e$ and $g$, also the constant $\delta$ is expected to be smaller than $1$ (the values used here are listed in table \ref{tab:model}).

\begin{figure}
  \centering
  \includegraphics[width=\textwidth]{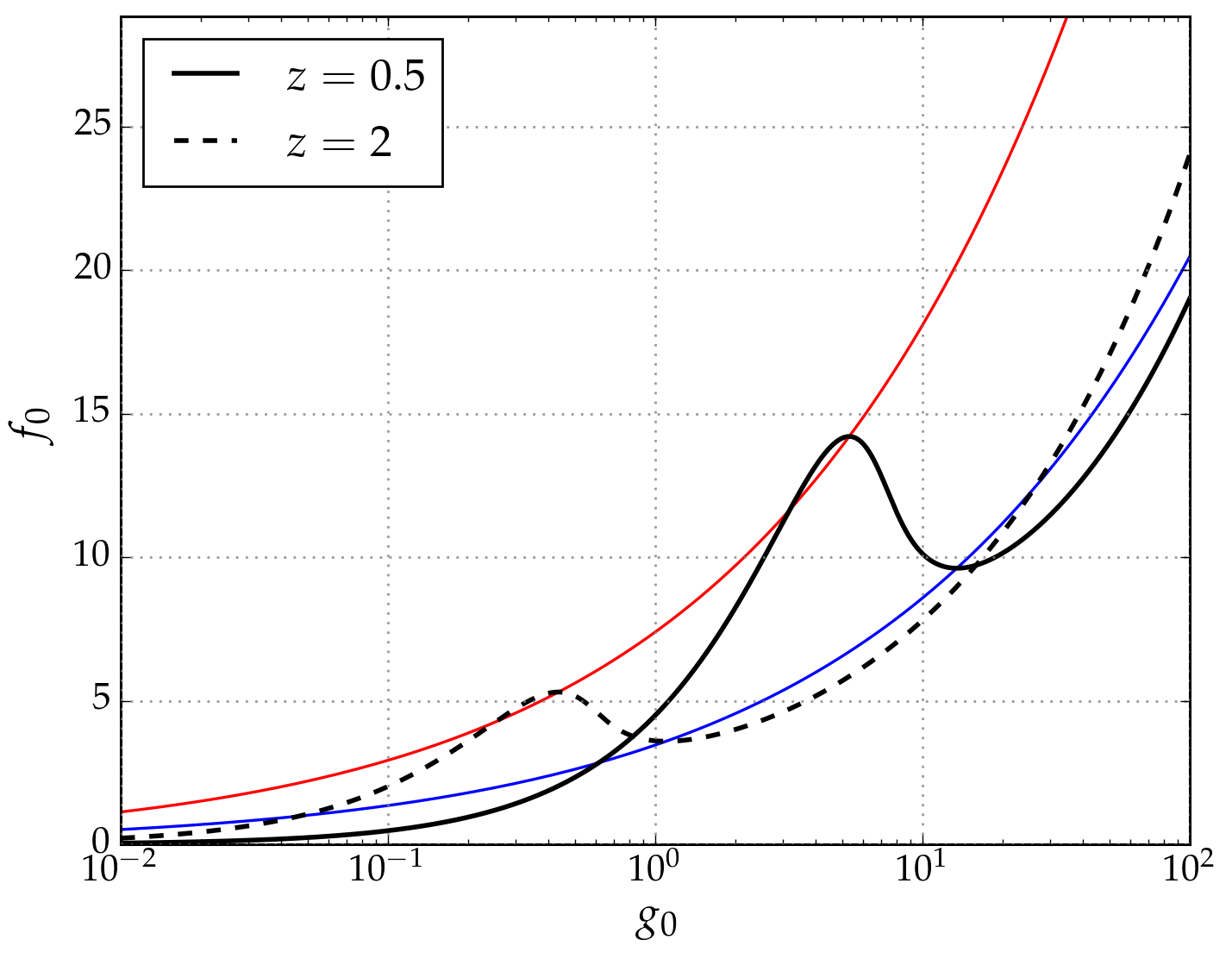}
  \caption{\label{fig:model}
    The non-dimensional equilibrium flux $f_0$ (from equations \ref{eq:eq_flux} and \ref{eq:eq_sol}) as a function of the non-dimensional gradient $g_0$ is shown using black lines for two values of the non-dimensional height above the bottom $z$ (see legend).
    The loci $(f_{max},g_{max})$ and $(f_{min},g_{min})$ are also shown in red and blue respectively.
    The results refer to the parameter values of table \ref{tab:model}.
  }
\end{figure}

\begin{figure}
  \centering
  \includegraphics[width=\textwidth]{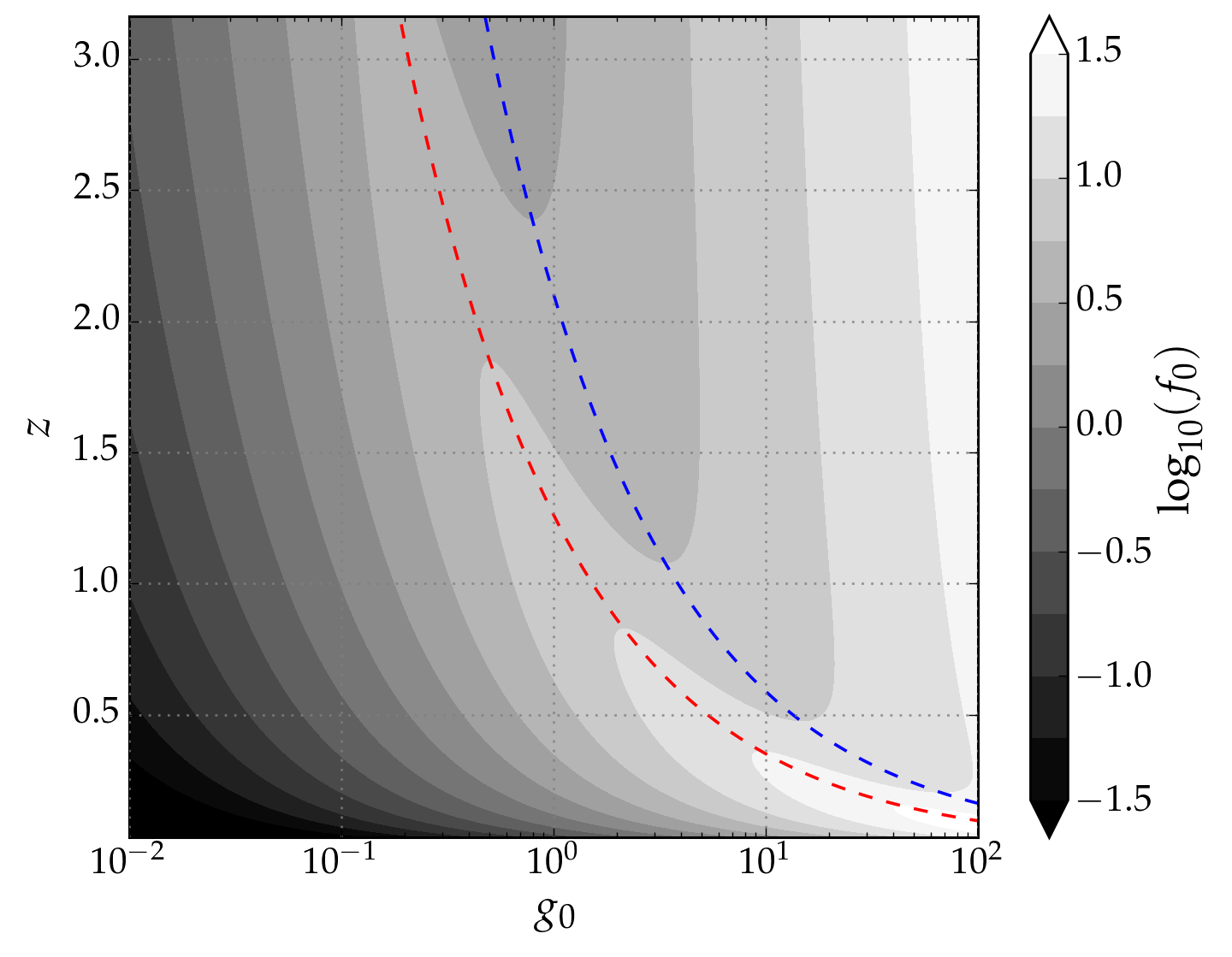}
  \caption{\label{fig:flux_model}
    Dependence of the non-dimensional equilibrium flux $f_0$ on $z$ and stratification $g_0$ in the model.
    The flux is shown with grey contours in logarithmic scale.
    The red (blue) dashed line marks the position of the local maximum (local minimum) of the flux predicted by the model ($\propto \log^2(\xi z)/z^2$).
    The results refer to the parameter values of table \ref{tab:model}.
  }
\end{figure}

In B98, the energy production term was defined as $\mathcal{P}=\alpha U^2 e^{1/2}l^{-1}$ with $U$ a constant velocity scale.
It was shown that $\mathcal{P}\propto e^{1/2}l^{-1}$ is essential for having an N-shaped flux-gradient relation.
In other words, $\mathcal{P}$ being inversely proportional to the eddy turnover time is essential for the model to produce layers of finite thickness.
Rather than by a stirring device moving at a constant velocity proportional to $U$, here the forcing is due to {the interaction of the internal wave field (mainly at the tidal frequency) with the sea-floor.
We hypothesise that the velocity profile in the region observed by the mooring ($z=5-105\m$) is approximated by the classical log-layer for constant-density fluids \citep{landau_fluid_1987}.
The hypothesis is not unreasonable, since the flow is highly turbulent (the Reynolds number is estimated at least $10^4$), but it neglects the effect of density stratification, internal wave reflection and rotation.
According to this hypothesis, the production term can be written as:
\begin{equation}
  \Pc = \alpha e^{1/2}l^{-1} u_*^2 \log^2\left(\frac{u_*}{\nu} z\right),
  \label{eq:forcing}
\end{equation}
with $u_*$ the shear velocity, assumed constant, and $\nu$ the molecular viscosity.
Note that, as in the classical theory, the factor $u_*/\nu$ in the argument of the logarithm is large if $u_*$ is of the order of the velocity fluctuations $u_*$ ($10^{-3}\m\s^{-1}$).
Our observations of velocity have insufficient resolution to evaluate how well a log-layer approximates the real velocity profile.
However, the small decrease of the mean velocity approaching the sea-floor (not shown) is not incompatible with \ref{eq:forcing}.

Non-dimensionalisation of the problem is carried out by taking:
\[
  z= \gamma^{-1/2} u_* N_0^{-1} \hat{z},\; g = N_0^2 \hat{g},\; e=u_*^2 \hat{e}\mbox{, and } l= \delta \gamma^{-1/2} u_* N_0^{-1} \hat{l},
\]
with hats indicating the non-dimensional quantities.

Combining the equations of the system \ref{eq:B98} in steady state ($\partial/\partial t\equiv 0$) with \ref{eq:mixing_length} and \ref{eq:forcing}, and applying the non-dimensionalisation, we obtain:
\begin{subeqnarray}
  \left[\frac{z e g}{\left(e+z^2 g\right)^{1/2}}\right]_{zz} = 0\\
  \beta r \gamma\left[\frac{z e e_z}{\left(e+z^2 g\right)^{1/2}}\right]_z - r \frac{z e g}{\left(e+z^2 g\right)^{1/2}} + \frac{\left(e+z^2 g\right)^{1/2}}{z} \left(\log^2\left(\xi z\right) - e \right) = 0,
\label{eq:model_ndim}
\end{subeqnarray}
where the hats for non-dimensional quantities have been dropped, $r=\delta^2/(\alpha\gamma)$ and $\xi=u_*^2/(N_0 \gamma^{1/2} \nu)$.
We could not derive an analytical solution of the system \ref{eq:model_ndim}.
Following B98, we rather seek equilibrium solutions of \ref{eq:model_ndim}-b neglecting the first term on the left hand side.
We thus solve an algebraic equation rather than a system of ordinary differential equations, by assuming that the first term on the left hand side of \ref{eq:model_ndim}-b is much smaller than 1 (i.e.~$\beta r \gamma\ll 1$).
From a physical point of view, this means that we neglect the vertical transport of kinetic energy in comparison to that of buoyancy.
This is a somewhat artificial assumption, used in order to make analytical progress, and which should be considered only as a first step in the analysis of \ref{eq:model_ndim}.
The investigation of the complete system \ref{eq:model_ndim} is the topic of further investigation, going beyond the scope of this work.

We thus solve the following equation for $e$:
\begin{equation}
  r z^2 e g - \left(e+z^2 g\right)\left(\log^2\left(\xi z\right)-e\right) = 0,
  \label{eq:nd_model}
\end{equation}
obtaining the (positive, physically meaningful) solution for the relationship between the equilibrium values $(g_0,e_0)$:
\begin{equation}
  \begin{split}
    e_0 = \frac{1}{2}&\left\{\log^2(\xi z)- z^2 g_0 (1+r) + \right. \\
    & \left. \left[\left(\log^2(\xi z)- z^2 g_0 (1+r)\right)^2 + 4 z^2 g_0 \log^2(\xi z)\right]^{1/2}\right\}.
  \end{split}
  \label{eq:eq_sol}
\end{equation}
This expression is very similar to equation 3.1 of B98, with additional terms $z^2$ and $\log^2(\xi z)$ due to our different definitions of $l$ and $\Pc$.
In the unstratified case, $g_0=0$ and $e_0=\log^2(\xi z)$, with the kinetic energy profile completely determined by the forcing.

The equilibrium value of the flux can be computed as:
\begin{equation}
  f_0 = l e_0^{1/2} g_0 = \frac{z e_0 g_0}{\left(e_0+z^2 g_0\right)^{1/2}}.
  \label{eq:eq_flux}
\end{equation}
Stationary points of the flux $f_0$ are found by solving $\partial f_0/\partial g_0 = 0$:
\begin{subeqnarray}
  g_{max} = \frac{2}{3 (1+r)^2} \left[2 (r-1) - \left(1-14 r+r^2\right)^{1/2}\right] \frac{\log^2(\xi z)}{z^2},\\
  g_{min} = \frac{2}{3 (1+r)^2} \left[2 (r-1) + \left(1-14 r+r^2\right)^{1/2}\right] \frac{\log^2(\xi z)}{z^2}.
  \label{eq:extrema}
\end{subeqnarray}
The corresponding expressions of the local maximum and minimum flux, $f_{max}$ and $f_{min}$ respectively, are complicated and unenlightening, and for this reason are only evaluated numerically.
As in B98, these stationary points of the flux are present only above a critical value, $r_c=7+4\sqrt{3}$.
Since we are interested in this model because of the presence of the stationary points in the flux--gradient relation, we focus our analysis on the case of $r$ well above $r_c$.
Different values of $r$ above the critical value give qualitatively similar results.
The non-dimensional flux as a function of the gradient is shown in figure \ref{fig:model} for two different values of $z$ and the parameters of table \ref{tab:model}.

Figure \ref{fig:flux_model} shows, in non-dimensional units, the flux $f_0$ as a function of both $g_0$ and $z$.
From figures \ref{fig:model} and \ref{fig:flux_model}, the main predictions of the model can be summarised:
\begin{enumerate}
\item the stationary points of the flux ($f_{max}$ and $f_{min}$) are shifted to smaller values of the gradient ($g_{max}$ and $g_{min}$) farther from the bottom;
\item the values of the gradient at the stationary points, $g_{max}$ and $g_{min}$, decrease with $z$ as $\log^2(\xi z)/z^2$.
\item the value of $f_{max}$ increases for decreasing $z$.
\item The flux tends to $f_0=e_0 g_0^{1/2}$ for $z^2 g_0\gg e_0$ (this asymptotic behaviour is the same as in the B98 model).
\end{enumerate}
With these predictions in mind, we now move to the estimation of the flux from observations.
}

\section{Methods}
\label{sec:methods}

\subsection{Mooring description and context}
\label{sec:dataset}

The data used in this work have been collected using a mooring deployed between spring and late summer 2013 on the slopes of Seamount Josephine (North-Eastern Atlantic Ocean, see table \ref{tab:mooring}).
The mooring is made of a $205\m$ long nylon-coated steel cable, attached approximately $5\m$ above a $500\kg$ ballast weight at the sea-floor, and to an elliptical buoy at the top.
A Teledyne/RDI 75 kHz acoustic Doppler current profiler (ADCP) was mounted in the top-buoy of the mooring, $210\m$ above the sea-floor.
The pressure and tilt sensors of the ADCP recorded movements smaller than $0.2\m$ in the vertical direction and $3\m$ in the horizontal one.

On the cable, 144 ``NIOZ4'' thermistors are taped at intervals of $0.7\m$.
The instruments are an evolution of the model described in \citet{van_haren_nioz3:_2009}.
The sampling rate of the thermistors is $1\Hz$, the precision of the temperature estimate is better than $1 \,\mathrm{mK}$, with noise level of approximately $5 \times 10^{-5}\dC$.
Each thermistor recorded a time series approximately $10^7\s$ long.

The top-mounted ADCP was pointed downwards, recording the horizontal velocity vector and echo intensity every $5 \m$ vertically, between $15$ and $185\m$ above the sea-floor.
The sampling interval of the ADCP was $600 \s$.

The calibration information for the raw temperature data is obtained from a calibration bath developed and operated at the Royal Netherlands Institute for Sea Research.
Compensation is made in the temperature data for the drift in the response of the thermistor electronics, which can be visible in some instruments over periods longer than a few weeks.

At the mooring location, semidiurnal tidal motions are dominant in both the velocity and temperature signals \citep{cimatoribus_comparison_2014,cimatoribus_temperature_2015}.
We divide the data in two subsets following \citet{cimatoribus_temperature_2015}, based on the phase of the tide.
The cooling and warming phases are defined by the sign of the time derivative of the vertically averaged temperature signal (negative and positive respectively), band-pass filtered at the semidiurnal frequency.
Velocities are mostly aligned to the isobath direction.
During the cooling phase, the cross-isobath component of velocity is preferentially upslope; vice-versa for the warming phase.
Mean velocities are larger during the cooling phase.
The vertical shear detected by the ADCP is small (due to limited vertical resolution).

\begin{figure}
  \centering
  \includegraphics[width=\textwidth]{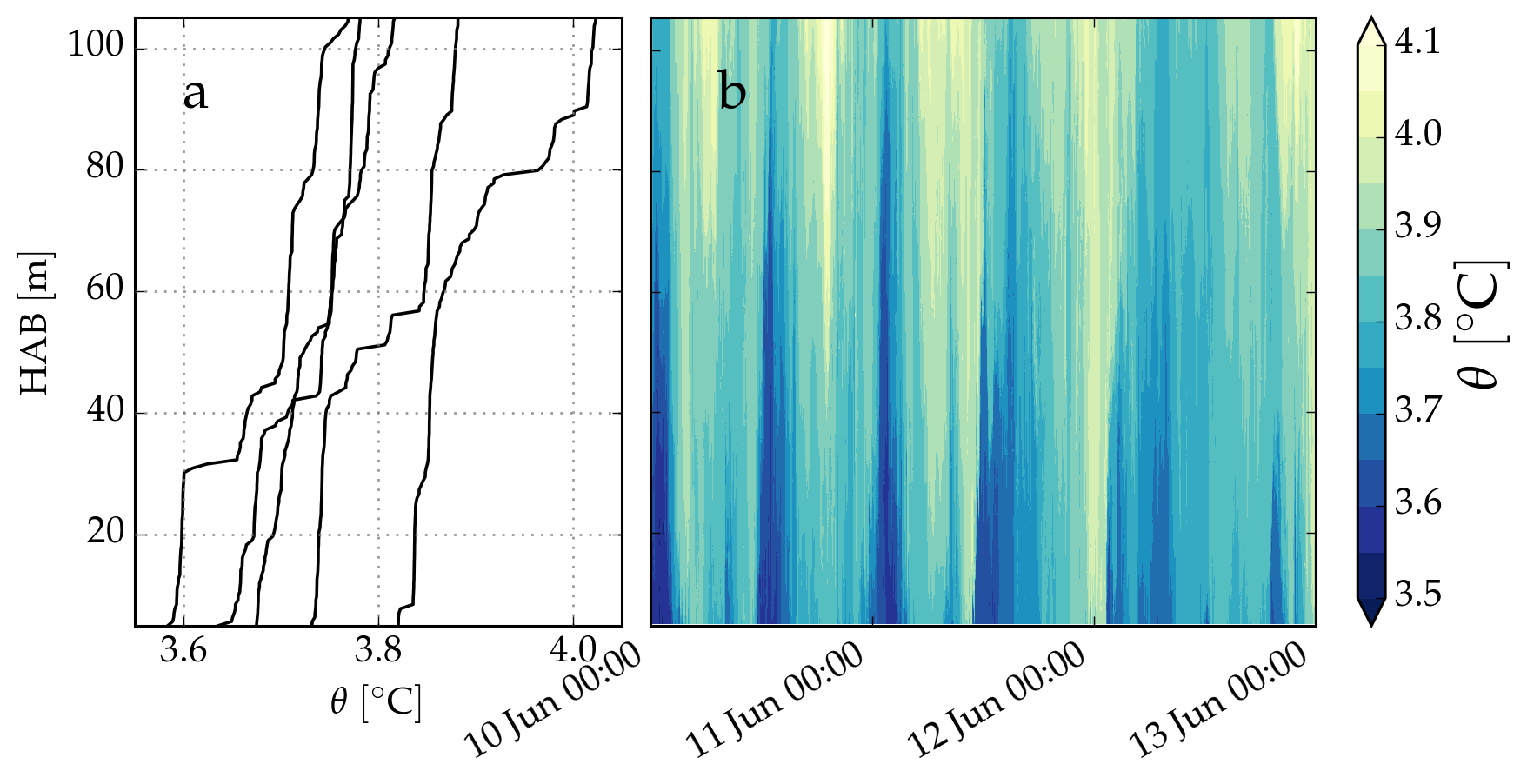}
  \caption{\label{fig:example}
  Panel a shows 5 vertical temperature profiles, randomly chosen in the dataset, reordered to remove any inversion.
  Panel b shows a contour plot with an example of the temperature data recorded by the mooring during three days, from 10 to 13 June 2013.
  }
\end{figure}

The temperature profiles are often organised in layers, with weakly stratified regions alternated with thinner sheets having strong stratification \citep{van_haren_where_2015,cimatoribus_temperature_2015}.
During the cooling phase, particularly strong stratification is often observed at the top of a cold temperature layer, more weakly stratified, extending to the bottom of the mooring ($5\m$ above the sea-floor).
See figure \ref{fig:example} for an example of the temperature data recorded at the mooring, in particular panel a, showing the ubiquitous presence of layers.

\subsection{Thorpe scale flux estimation method}
\label{sec:thorpe}

At the time of deployment and recovery, two shipborne Conductivity Temperature Depth (CTD) surveys were performed, for comparison with the results from the thermistors and to check the temperature--density relationship.
As discussed in more detail in \cite{cimatoribus_comparison_2014}, the low noise level of the temperature data in combination with the tight and linear temperature--density relationship observed in the CTD data, support the use of temperature as a proxy for buoyancy.
This enables the estimation of overturning scales, providing in turn an estimate of turbulent diffusivity.

The Thorpe scale ($l_T$) is the root mean square (rms) of the displacements necessary to adiabatically reorder a buoyancy profile, to remove all buoyancy inversions  \citep{thorpe_turbulence_1977}.
$l_T$ is a measure of the overturning caused by turbulent eddies and can be related to the Ozmidov scale $l_O$, the scale at which buoyancy and inertial forces balance in a stratified flow \citep{thorpe_turbulence_1977,dillon_vertical_1982}.
Using the empirical relation of \cite{dillon_vertical_1982}, the Thorpe scale can provide an estimate of vertical diffusivity and flux:
\begin{equation}
  K_\theta^{T} = 0.128\, l_T^2 N,
  \label{eq:diff_thorpe}
\end{equation}
where $\theta$ is the potential temperature, $N$ is the buoyancy frequency and $0.128$ is an empirical constant.
The superscript is a reminder of the method used for obtaining the estimate.
Multiplication of diffusivity from equation \ref{eq:diff_thorpe} with the background temperature gradient provides a Thorpe scale-based estimate of the flux, which will be indicated by $\fT$.
The flux estimate $\fT$ is the diffusive one, rather than the total which includes the advective one \citep[see e.g.][]{scotti_biases_2015}.
 While experiments show that the advective flux may dominate the total \citep{martin_layering_2006}, the irreversible flux is the component actually changing the available potential energy of the system irreversibly.

 Since $K_\theta^T$ would tend to the molecular value for vanishing $l_T$, molecular diffusivity may produce an N-shaped flux-gradient relation in highly stable conditions.
 However, this is not observed here, since the estimates of the flux obtained from \ref{eq:diff_thorpe} are at least 3 orders of magnitude higher than the molecular values.
 Equation \ref{eq:diff_thorpe} is based on the assumption of a constant mixing efficiency, whose validity is a subject of active discussion \citep[see][and references therein]{mater_quest_2014}, and which is used here being the simplest option and the most common one in the oceanographic context.

Furthermore, equation \ref{eq:diff_thorpe} can be derived only for the case of fully developed turbulence outside boundary layers.
Turbulence is indeed strong in this area with internal waves constantly breaking and producing overturns throughout the depth range spanned by the mooring.
We expect that the presence of the sea-floor biases the Thorpe scale-based estimates of diffusivity and flux, due to the introduction of an external length-scale (the distance from the boundary), as suggested by \citet{thorpe_turbulence_1977}.
This is checked by comparing estimates of $\fT$ computed excluding the deeper parts of the mooring in section \ref{sec:results}.

We apply the Thorpe reordering approach in combination with equation \ref{eq:diff_thorpe} to our data, averaged in $200\s$ long, independent time-windows, obtaining a time series of vertically averaged estimates of diffusivity.
No isolated overturns can be identified in this (highly turbulent) data set, and thus the vertical averaging is performed over the whole mooring depth (in some cases, excluding the deepest $25\m$ of the mooring, see section \ref{sec:results}).

\subsection{Winters and D'Asaro flux estimation method}
\label{sec:winters-d'asaro}

The Thorpe scale-based method for estimating the temperature flux cannot resolve the vertical details of the turbulent fluxes, since it is based on the rms displacement in the whole profile.
In this context, the method suggested by \citet{winters_diascalar_1996} provides a framework to study the vertical dependence of turbulent fluxes.
We outline it here.
The advection-diffusion equation for a scalar (in this case the potential temperature $\theta$) in a fluid flow is:
\begin{equation}
  \frac{\partial \theta}{\partial t} + \mathbf{u}\cdot\mathbf{\nabla} \theta = \kappa \nabla^2 \theta,
  \label{eq:adv-diff}
\end{equation}
where $t$ is time, $\mathbf{u}$ is the velocity vector and $\kappa$ is the coefficient of molecular diffusivity.
\citet{winters_diascalar_1996} considered the purely diffusive flux $\kappa \nabla \theta$, i.e.~the irreversible flux, averaged over an isothermal surface $S$:
\begin{equation}
  \fW = \frac{1}{A}\int_S \kappa \nabla \theta \cdot \hat{n}\; \dd S.
  \label{eq:diff-flux}
\end{equation}
In equation \ref{eq:diff-flux}, $\hat{n}$ is the unit vector normal to the isothermal surface $S$ (pointing in the direction of increasing $\theta$).
The flux is normalised by $A$, the cross-sectional area of the surface we are considering, rather than by $A_S$, the surface area of $S$ ($A_S\ge A$).
\citet{winters_diascalar_1996} showed that $\fW$ can be expressed as:
\begin{equation}
  \fW = \kappa\frac{\dd \zs}{\dd \theta}\left<\left|\nabla \theta\right |^2\right>,
  \label{eq:diff-flux-zstar}
\end{equation}
with brackets indicating isothermal averaging.
The coordinate $z^*$ is defined by:
\begin{equation}
  \zs (\mathbf{x},t) = \frac{1}{A}\int_V \mathcal{H}(\theta(\mathbf{x},t)-\theta(\mathbf{x}',t))\; \dd^3 \mathbf{x}',
  \label{eq:zstar}
\end{equation}
with $V$ the volume containing the fluid, $\mathbf{x}$ the position vector and $\mathcal{H}$ the Heaviside function.
Equation \ref{eq:zstar} defines a reordering procedure, mapping the three-dimensional volume occupied by the fluid onto the one-dimensional coordinate $\zs$.
The coordinate $\zs$ has a unique value for all points on an isothermal surface, with $\zs(\theta_2)>\zs(\theta_1)$ for $\theta_2>\theta_1$.

Note that the Osborn-Cox method provides an estimate of the average advective flux rather than of the diffusive flux \ref{eq:diff-flux}; these two fluxes are not necessarily equal \citep[see][]{winters_diascalar_1996}.
While the flux estimates according to the Osborn-Cox method have been computed, their correlation coefficient with the Thorpe scales-based estimates is lower ($0.6$ versus $0.8$).
This is not surprising, since both Winters-D'Asaro and Thorpe methods provide an estimate of the diffusive flux, while the Osborn-Cox method provides an estimate of the advective flux.

Equation \ref{eq:diff-flux-zstar} is derived without any assumption on the flow, and is thus correct for scalars satisfying \ref{eq:adv-diff} in incompressible flows.
However, its use requires the complete knowledge of the three-dimensional temperature field, and is thus not directly applicable in an oceanographic context, because observations are usually one-dimensional profiles.
\citet{winters_diascalar_1996} suggested a method to approximate \ref{eq:diff-flux-zstar} if a series of one-dimensional profiles is available.
We adapt the method of \citet{winters_diascalar_1996} to our particular data set in the following way.

Each profile provides a series of potential temperature measurement $\theta_i(z_j)$ at equally spaced vertical positions $z_j$ with $j=1,2,\ldots M$ ($z_n>z_m$ for $n>m$), and at equally spaced times $i$.
At $z_j$, the temperature gradient $\left|\nabla\theta\right|_i(z_j)$ is estimated, using the Taylor's hypothesis for obtaining its horizontal component (see appendix \ref{sec:appendixA}).
$P$ consecutive profiles in time are combined to provide an estimate of the flux.
The procedure is as follows \citep{winters_diascalar_1996}:
\begin{enumerate}
\item Each profile, of both $\theta$ and its gradient, is sorted in ascending order of potential temperature. The original profiles $\theta_i(z_j)$ and $\left|\nabla\theta\right|_i(z_j)$ thus become $\theta_i(z_j^T)$ and $\left|\nabla\theta\right|_i(z_j^T)$ where $z_j^T$ is the equally spaced depth coordinate obtained from the ``Thorpe reordering'' method of \citet{thorpe_turbulence_1977}.
\item Each profile $\theta_i(z_j^T)$ and $\left|\nabla\theta\right|_i(z_j^T)$, having constant sampling rate, is interpolated onto a set of equally spaced values of potential temperature $\theta_j$, obtaining the new profiles  $\theta_i(\theta_j)$ and $\left|\nabla\theta\right|_i(\theta_j)$.
The sampling rate of these new profiles is $1\times 10^{-3}\dC$, a safe upper limit of the instruments precision.
\item Isothermal averages are formed:
  \begin{subeqnarray}
     \left<\left|\nabla\theta\right|^2\right>^P(\theta_j) = \frac{1}{P}\sum_{i=1}^P\left|\nabla\theta\right|^2_i(\theta_j), \\
     \left<z^T\right>^P(\theta_j) = \frac{1}{P}\sum_{i=1}^Pz^T_i(\theta_j).
     \label{eq:averages}
   \end{subeqnarray}
The hypothesis is that the $P$-averages of equation \ref{eq:averages} provide a good estimate of the local isoscalar average of the temperature gradient, and of the  $\zs$ coordinate respectively.
\item Using the profile $\left<z^T\right>^P(\theta_j)$, an estimate $(\dd z^T/\dd\theta)^P(\theta_j)$ of the background gradient is computed using centred discrete derivatives.
\item The estimate of the flux is finally provided by rewriting \ref{eq:diff-flux-zstar} as:
  \begin{equation}
    \fW(\theta_j) = \kappa\left(\frac{\dd z^T}{\dd \theta}\right)^P(\theta_j) \left<\left|\nabla\theta\right|^2\right>^P(\theta_j).
    \label{eq:N-diff-flux}
  \end{equation}
  The flux is positive downwards.
\item The estimate of the flux of equation \ref{eq:N-diff-flux} is interpolated back on the depth coordinate $z_j$.
\end{enumerate}

The results presented here use $P=200$, and thus the quantities are averages over $200\s$ of data.
This time-scale is chosen for being the longest one still shorter than all fast gravity wave periods ($2\mathrm{\pi}/N$).
In other words, during $200\s$, the measured temperature changes due to turbulence and horizontal advection, rather than due to internal waves.
Since the mean velocity is of the order of $0.1\m\s^{-1}$, the horizontal displacement during this time is of the order of $10\m$.
The choice $P=200$ thus avoids averaging regions with widely different dynamics, while reducing the uncertainty of the estimates (a turbulent overturn in the data can easily span tens of meters in the vertical).
The use of different values of $P$ ($\pm 100$) does not change the results significantly, except for an obvious increase of noise for lower $P$'s and a decrease of the dynamic range for higher $P$'s.

An essential element in the method is the interpolation from the depth coordinate to the temperature coordinate, which provides a consistent way to estimate the background gradient, on an isotherm rather than on an isobath, and thus provides a consistent definition of the flux at each depth (after interpolating back onto the depth coordinate).

It is important to note that the estimate of the temperature gradient $\nabla\theta$, obtained from the measurements, is biased low.
This is a well known issue, already noted by \citet{thorpe_turbulence_1977}, related to the fact that the instruments do not resolve the length scales of molecular diffusion \cite[see][for further details]{cimatoribus_temperature_2015}.
For a more general discussion of the wavenumber spectrum of scalar gradients, see \citet{warhaft_passive_2000}.
To estimate the impact of this underestimation, we compare the values of $\fW$ from equation \ref{eq:N-diff-flux} to those obtained using the classical Thorpe method, considered for the moment the ground truth.
In the next section, we will show that the two estimates are correlated; a linear fit in the log-log space thus provides an empirical way to compensate for the underestimation of the gradient.
While a completely independent, fully resolved measurement would ideally be the only term of comparison, we argue that this approach is a plausible, albeit imperfect, one.

\section{Results}
\label{sec:results}

\subsection{Comparison between the two flux estimates}
\label{sec:comparison}

Since the Thorpe scale-based estimate is a vertical average, we consider the vertical average of $\fW$ as well.
As already discussed, $\fT$ is expected to provide a good estimate of the average flux far from the boundary alone.
Close to the sea-floor, the Thorpe scale is limited by the distance from the sea-floor itself rather than by the buoyancy gradient, as hypothesised in the derivation of equation \ref{eq:diff_thorpe}.
On the other hand, the estimate \ref{eq:N-diff-flux} does not rely on any such hypothesis, and is thus expected to be valid everywhere, within the limits due to the under-resolution of temperature gradients discussed above.

\begin{figure}
  \centering
  \includegraphics[width=0.7\textwidth]{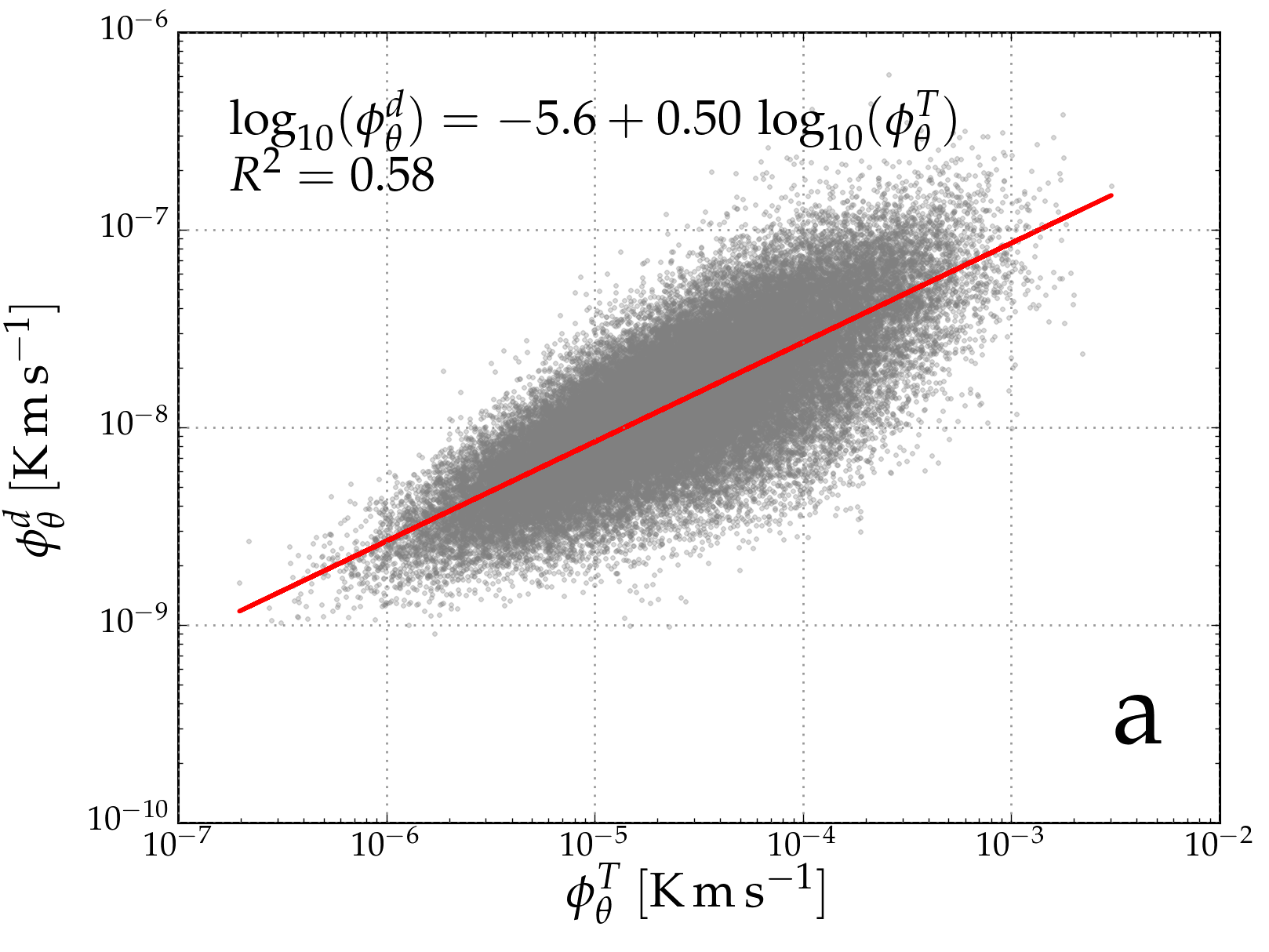}
  \includegraphics[width=0.7\textwidth]{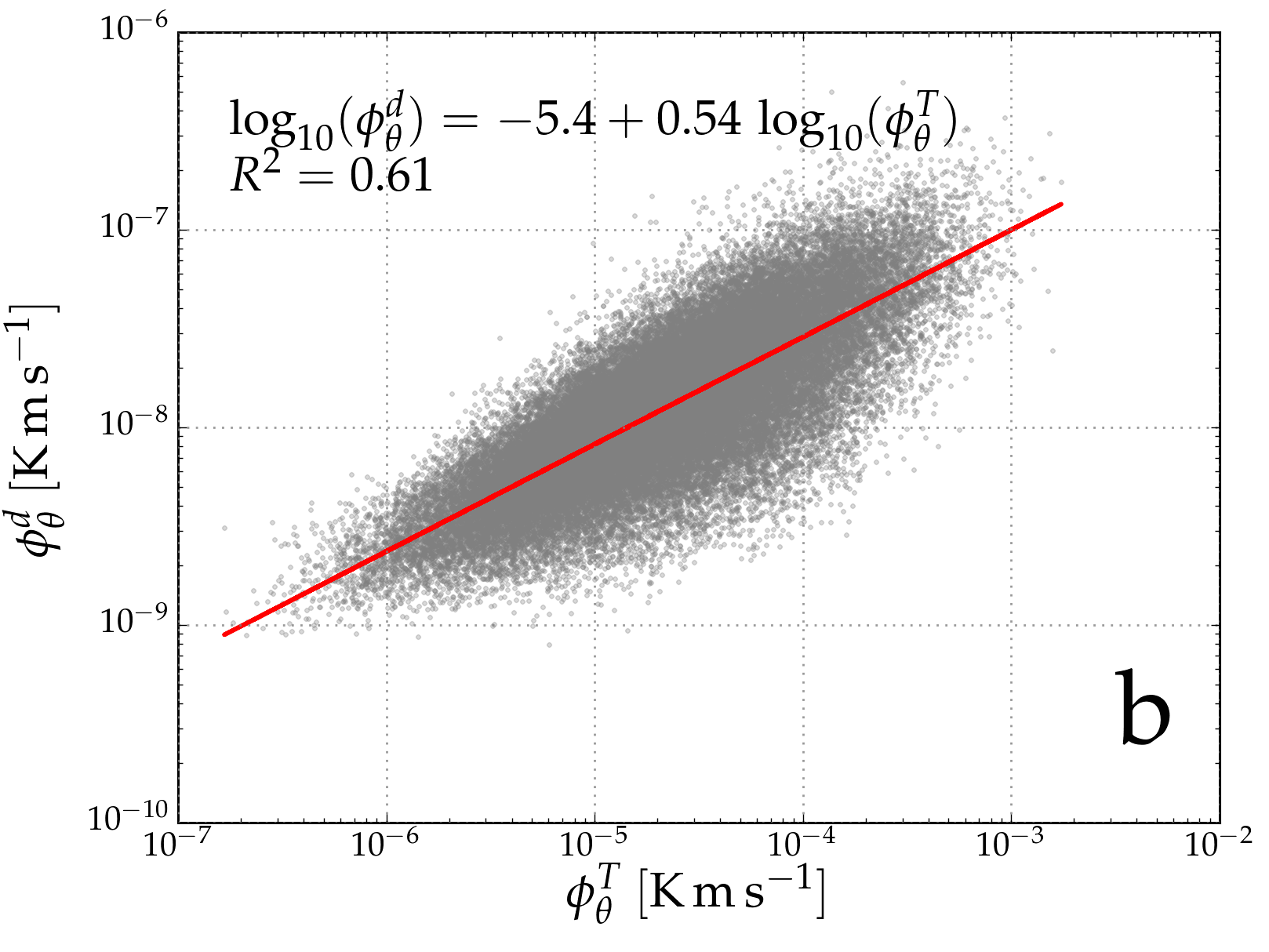}
  \caption{\label{fig:flux_comparison}
    Comparison of the two flux estimates using the Thorpe scale method ($\fT$, section \ref{sec:thorpe}) and the Winters-D'Asaro method ($\fW$, section \ref{sec:winters-d'asaro}).
    Panel a shows the comparison of the flux estimates vertically averaged over the entire mooring.
    Panel b shows the comparison of the flux estimates averaged excluding the lower $25\m$ of the mooring.
    The red line in each panel is a linear fit in the log-log space, the result of the fit is shown.
  }
\end{figure}

Figure \ref{fig:flux_comparison}-a shows the comparison of the two estimates in log-log space, vertically averaged over the whole mooring.
Interpolation in time is performed to align the two time series with a time step of $200\s$.
A linear fit of the two estimates in the log-log space gives $\log_{10}(\fW)=-5.6+0.5\,\log_{10}(\fT)$.
The coefficient of determination of this fit is $R^2=0.58$.
The two estimates are thus correlated, with $\fW$ approximately proportional to $(\fT)^{1/2}$, even if there is significant spread in the data.
The spread may be due to the intrinsic limitations of the estimation methods (in particular for $\fT$), or may be connected to the fact that we do not resolve the entire three-dimensional turbulent field with our measurements.
The large underestimation of $\fT$ by $\fW$ is the effect of limited resolution, as discussed above.

With the aim of assessing if the closeness to the boundary has a significant effect on the flux estimates, we show in figure \ref{fig:flux_comparison}-b the two flux estimates, vertically averaged excluding the bottom $25\m$ of the mooring.
In the case of the Thorpe estimate, not defined at each depth, we compute a value of the diffusivity only using the Thorpe displacements in the top $75\m$ of the mooring.
The results are very similar to those using the whole depth range, and the increase in the coefficient of determination is marginal ($R^2=0.61$).
We thus suggest that, at least in this data set, the effect of the solid boundary on the Thorpe scale flux estimates is small.
This result will be discussed in more detail in section \ref{sec:conclusions}.
We must note, however, that our measurements do not extend below an HAB of approximately $5\m$.

Assuming that the Thorpe scale flux estimate is correct (on average), we correct the underestimated $\fW$ using the results of the fit in figure \ref{fig:flux_comparison}-a.
This approach can be compared to the common practice of fitting the Batchelor wavenumber spectrum on underresolved temperature measurements to obtain an estimate of the temperature variance dissipation rates \citep{ruddick_maximum_2000}.
By necessity, we are also assuming that the correction can be applied in the same way for each estimate $\fW$ (at all times and depths).
This ``corrected'' value of the flux will be denoted by $\fWc$.

\subsection{Dependence of $\fWc$ on depth and stratification}
\label{sec:flux-dependence}

As discussed in section \ref{sec:winters-d'asaro}, the method proposed by \citet{winters_diascalar_1996} for estimating the temperature (or other scalar) flux enables a consistent definition of the flux and of the background gradient at each depth.
In this section, we fully exploit this advantage, studying the dependence of the flux on the local stratification and on the HAB.

\begin{figure}
  \centering
  \includegraphics[width=0.7\textwidth]{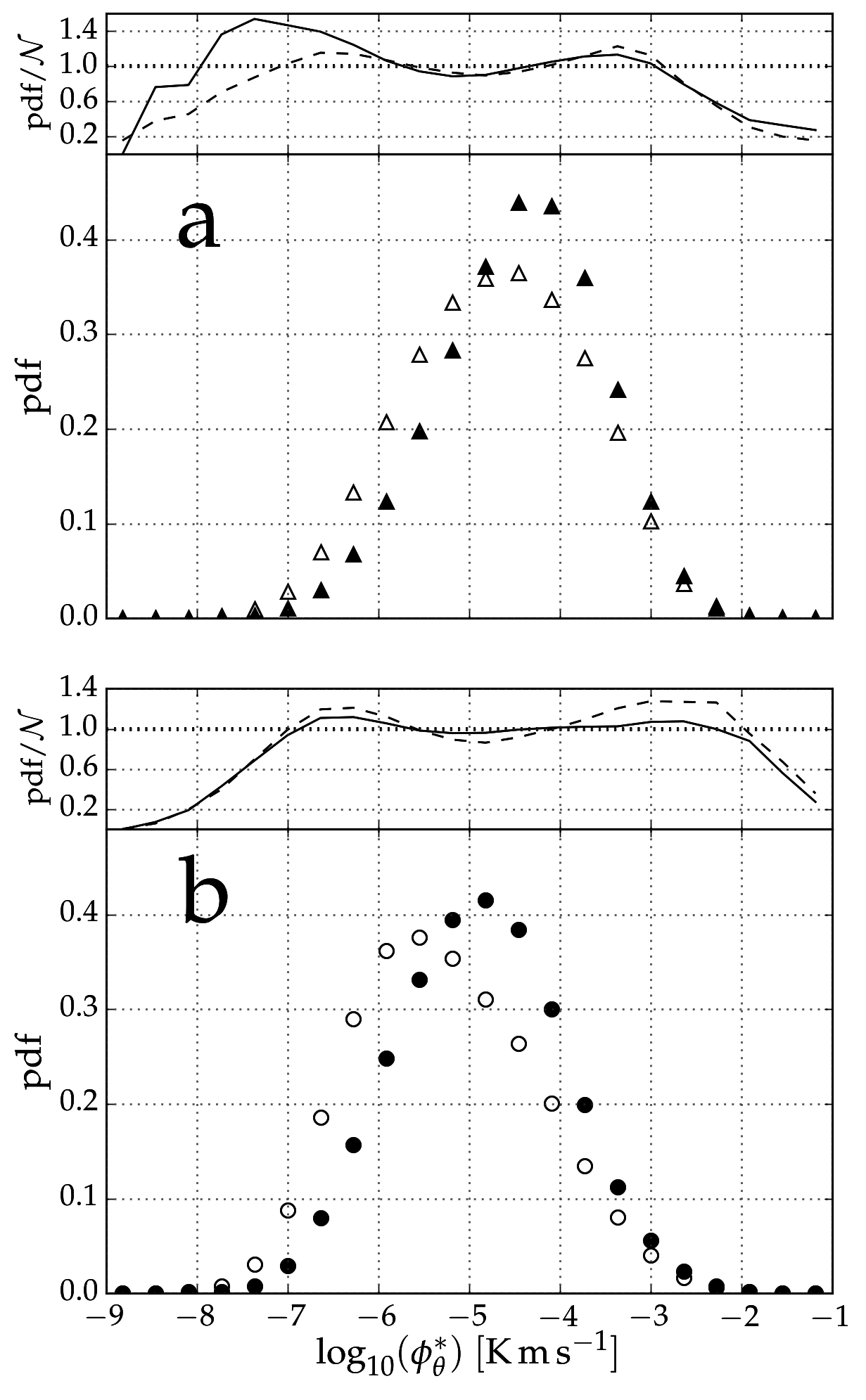}
  \caption{\label{fig:flux_pdf}
    Pdf's of the logarithm of the flux ($\log_{10}(\fWc)$), compensated for underestimation of the temperature gradient (see text).
    Panel a shows the results during the cooling phase of the tide, panel b those during the warming phase.
    The insets above each panel show the ratio of the pdf with a normal distribution, demonstrating that the flux has an approximately log-normal distribution.
    Filled symbols refer to data from the lower half of the mooring (continuous line in the inset), empty symbols to data from the upper half of the mooring (dashed line in the inset).
    The horizontal scale is logarithmic.
  }
\end{figure}

The probability density functions (pdf's) of the flux estimates $\fWc$ are shown in figure \ref{fig:flux_pdf}, computed separately for the cooling and warming tidal phases (panels a and b respectively, see section \ref{sec:dataset} for the definition of the phases).
During each tidal phase, two pdf's are computed, one for data from the upper half of the mooring (empty symbols) and one for data from the lower half (filled symbols).
The insets in the figure show the ratio between the empirical distribution and a normal distribution having the mean and the standard deviation of the log-transformed data.
The pdf's all show an approximately log-normal behaviour close to the mean, similarly to what is observed for dissipation, both in the laboratory and in the ocean \citep[see e.g.][]{baker_sampling_1987,warhaft_passive_2000}.
Deviations from log-normality are strong only in the tails, where undersampling leads to shorter tails in the empirical distribution in comparison to the theoretical distribution.
Deviations from log-normality are even smaller for the pdf of the entire data set (not shown).
The effect of noise, leading to fatter left tails, is hardly visible only in the cooling phase close to the sea-floor, testifying to the extremely low noise-level of the instruments.
Values of the flux higher by almost one order of magnitude are observed during the cooling phase; higher values of the flux are also observed in the lower half of the mooring.

\begin{figure}
  \centering
  \includegraphics[width=\textwidth]{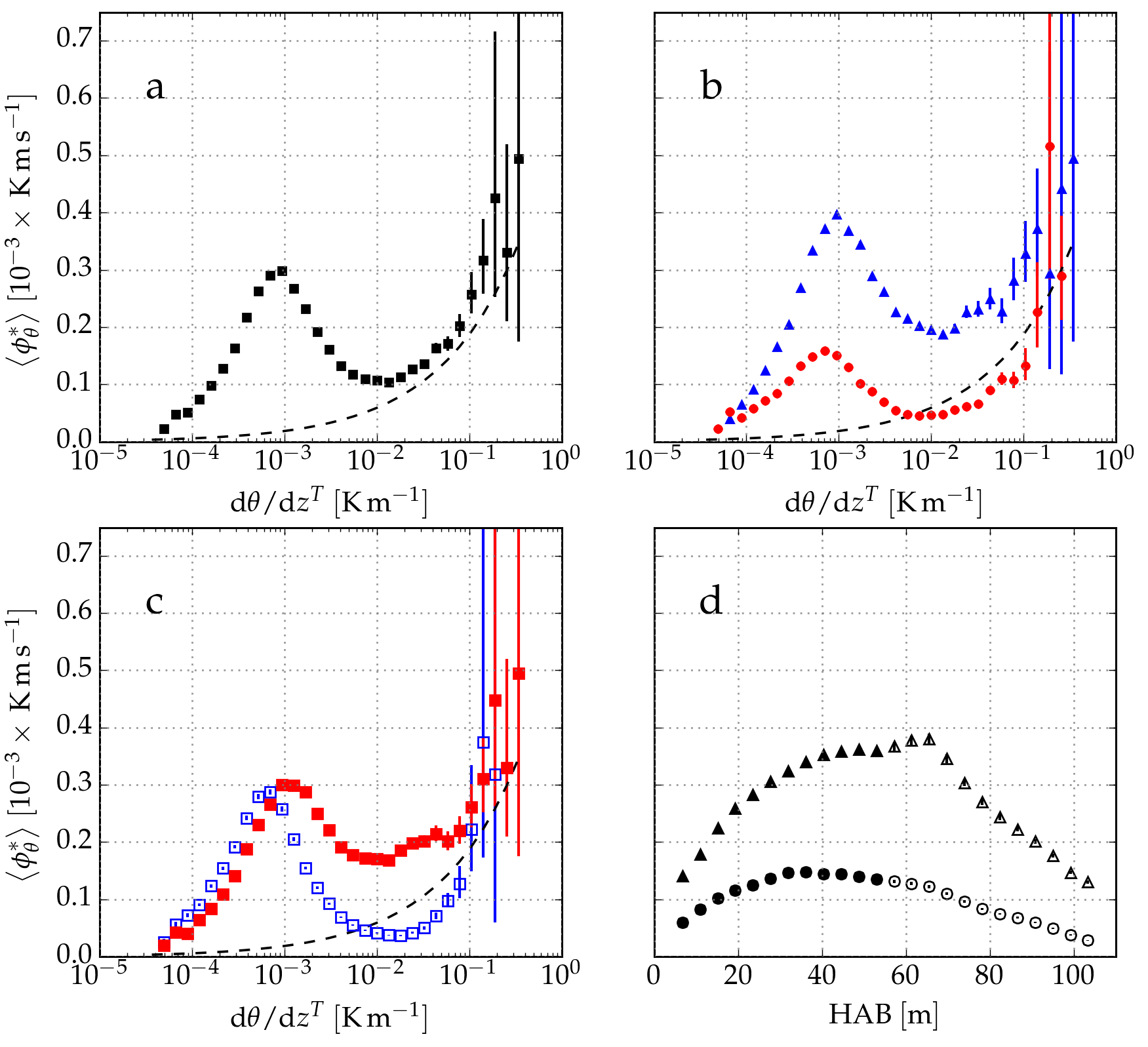}
  \caption{\label{fig:flux_scatter}
    Temperature flux $\fWc$ as a function of background stratification $\dd \theta/\dd z^T$ (panels a, b and c) and as a function of the height above the bottom (panel d).
    Error bars, visible only for large values of $\dd \theta/\dd z^T$, are $95\%$ confidence intervals, computed using the Cox's method described and evaluated in \citet{zhou_confidence_1997}.
    Panel a shows averages of all the flux estimates; panel b shows the estimates averaged separately during the cooling (blue triangles) and warming (red circles) tidal phase; panel c shows the estimates averaged separately in the lower half (red filled symbols) and upper half (blue empty symbols) of the mooring; panel d shows the depth dependence of the flux, estimated separately for the cooling (triangles) and warming (circles) tidal phase.
    The dashed line in panels a, b and c is a guide for the eye $\propto (\dd \theta/\dd z^T)^{1/2}$ (see text).
  }
\end{figure}

In figure \ref{fig:flux_scatter}-a, estimates of $\fWc$ are shown, averaged in 32 bins of the background temperature gradient $\dd \theta/\dd z^T$ (equally spaced in log space).
Since $\fWc$ is log-normally distributed, the mean is estimated from $\exp(\mu+\sigma^2/2)$, with $\mu$ and $\sigma^2$ the mean and variance of the log-transformed data respectively, as motivated e.g.~in \citet{baker_sampling_1987}.
While this procedure is not essential, it improves the results in particular at the highest temperature gradients, for which less data points are available.
The average flux has a tight, non-linear dependence on stratification, with a local maximum for $\dd\theta/\dd z^T\approx 10^{-3}\dC\m^{-1}$, a local minimum at $\dd\theta/\dd z^T\approx 10^{-2}\dC\m^{-1}$, and a new increase for higher values of the stratification.
It is impressive to see how closely this curve resembles the flux-gradient relationship hypothesised by \citet{phillips_turbulence_1972}, \citet{posmentier_generation_1977} and B98, and predicted by our model (section \ref{sec:theory}, figure \ref{fig:model}).
As expected, the flux also grows as the square root of the gradient for high values of the gradient (dashed line).
We stress that such a smooth curve can be obtained only after sufficient averaging (i.e.~using a large data set as the one here).
We would expect such a relation to hold locally as well, had we sufficient knowledge of the flow to extract it reliably.

Figure \ref{fig:flux_scatter}-a could suggest that the diffusive flux in a turbulent flow can be universally described by a function of stratification alone.
However, this is obviously not the case, as is clear from a more detailed analysis of the data set.
Figure \ref{fig:flux_scatter}-b shows the mean flux averaged separately during the cooling and warming phases.
Stronger fluxes take place during the cooling phase (blue triangles), with the position of the stationary points virtually the same as the ones of the warming phase (red circles).

Figure \ref{fig:flux_scatter}-c shows the mean flux averaged separately in the lower half (filled red squares) and in the upper half of the mooring (empty blue squares).
The curve from the data closer to the bottom has a local maximum at a higher value of stratification, and a less pronounced local minimum.
The comparison of panels a, b and c suggests that the flux is a function of stratification alone for very strong stratification, at least within the large error bars of the estimates therein.
For strong stratification ($\dd\theta/\dd z^T>10^{-1}\mathrm{K}\m^{-1}$), the flux increases as $(\dd\theta/\dd z^T)^{1/2}$, as predicted by the theory in section \ref{sec:theory}.
A small bump is visible in flux-gradient relation at $\dd\theta/\dd z^T\approx 5\times 10^{-2}\dC\m^{-1}$ in panels b and c (red symbols, lower half of the mooring).
This barely visible feature is robust to changes in the number of profiles used for averaging ($P$ of the Winters-D'Asaro estimates).

Figure \ref{fig:flux_scatter}-d shows the dependence of the flux on the HAB, as obtained from averaging the data in 24 equally spaced depth bins, separately during the two tidal phases (triangles for cooling phase, circles for warming phase).
The essential new information here is that the flux has a maximum at approximately $30$-$40\m$ above the bottom, which must be linked to horizontal advection if the profile is stationary.
During each tidal phase, the maximum averaged flux value is approximately a factor 2 of the minimum.

The dependence of $\fWc$ on the gradient and HAB is summarised as a contour plot in figure \ref{fig:flux_obs}, showing also the estimated positions of the local maxima and minima of the flux.
This figure will be compared to figure \ref{fig:flux_model} in the next section.

\section{Discussion and conclusions}
\label{sec:conclusions}

We presented observations of the temperature flux in the deep ocean, above a sloping sea-floor.
The observations provide detail on the dependence of the flux on the local vertical temperature gradient, and on the height above the sea-floor.
They support the predictions of the model for the flux-gradient-HAB relation discussed in section \ref{sec:theory}.
This is surprising, in particular considering that we solved a highly simplified version of the model, assuming steady state and negligible diffusion of turbulent kinetic energy.
It is however important to stress once more that averaging is essential here: only using a high-resolution, long data set it is possible to extract the smooth curves of figure \ref{fig:flux_scatter}, which are otherwise obscured by the huge variability of the turbulent quantities.
Furthermore, the use of the method outlined by \citet{winters_diascalar_1996} is essential for correctly disentangling the flux-gradient-HAB relation as well.
While a more spatially complete observation of the turbulent flow is expected to reduce the need for averaging, and to bring different estimation methods to convergence, we are very far from this situation, both in the ocean and, possibly to a somewhat lesser extent, also in the laboratory.

\begin{figure}
  \centering
  \includegraphics[width=\textwidth]{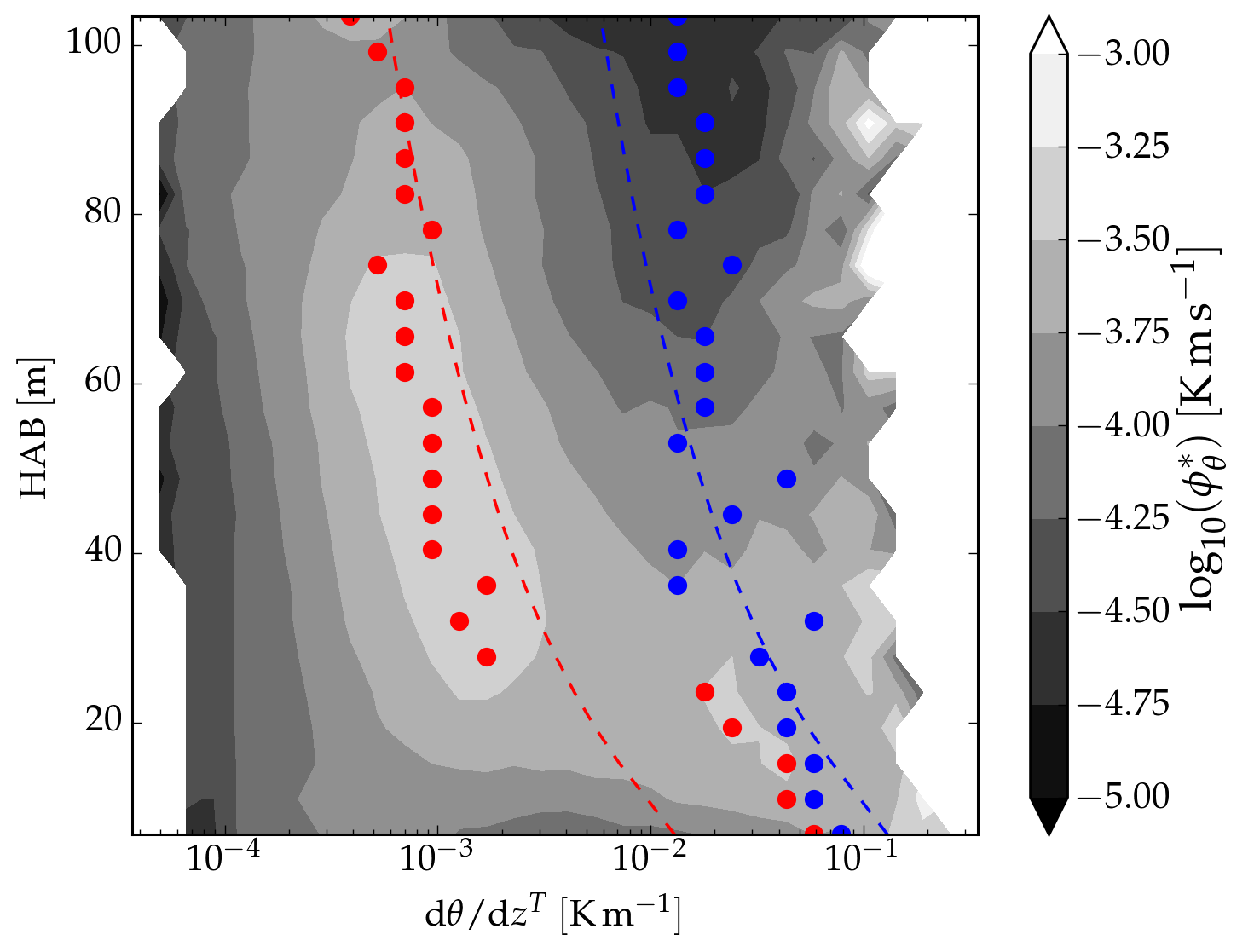}
  \caption{\label{fig:flux_obs}
    Dependence of the $\fWc$ on stratification and HAB, shown in grey contours with a  logarithmic scale.
    Red (blue) dots mark the positions of the local maximum (minimum) of the flux.
    The red and blue dashed lines show the positions of the stationary points predicted by the model (see text).
  }
\end{figure}

The agreement between model and observations supports the mixing-length hypothesis, in particular when considering the similar patterns observed in the model and in the observations as a function of the HAB (compare figures \ref{fig:flux_model} and \ref{fig:flux_obs}).
The agreement is also supported by the fact that model and observations span a similar range of scales (the model results being presented in non-dimensional form) in terms of gradient, HAB and flux.
Furthermore, the results strongly suggest that the solid boundary limits the mixing length.
\citet{thorpe_turbulence_1977} argued that the Thorpe scale could not be used as a proxy for turbulence dissipation close to a boundary, since the distance from the boundary, rather than stratification, limits the overturn scale therein.
While our results support the latter argument, they also suggest that the mixing length is limited by the wall in a way similar to the Thorpe scale, in particular considering the model and the comparison between $\fW$ and $\fT$.
This in turn supports the use of the Thorpe scale close to the boundary for obtaining an order-of-magnitude estimate of turbulence parameters, at least as long as stratification is present and averaging is performed.

{It is interesting to consider the positions of the stationary flux points in the model and in the observations.
This is done in figure \ref{fig:flux_obs}, reporting as red (blue) dots the position of the maxima (minima) of the flux-gradient relation estimated from the observations, and with similarly coloured dashed lines those predicted by the model.
The stationary points of the model, according to equation \ref{eq:extrema}, are loci where the gradient is proportional to $\log^2(\eta z)/z^2$, using $\eta=0.5\m^{-1}$.
The model and the observations roughly agree.

It should be noted that the value of $\eta$ used in figure \ref{fig:flux_obs} is much smaller than the one computed from the values in table \ref{tab:model}.
The classical theory, which we followed, states that $\eta=u_*/\nu$, with $u_*$ of the same order of magnitude as the velocity fluctuations.
If we use a larger value of $\eta$, e.g.~$10^3\m^{-1}$ as implied by the values in table \ref{tab:model}, a much worse agreement with the observations is obtained.
This is confirmed by fitting a function $\dd \theta / \dd z^T \propto \log^2(\eta z)/z^2$ to the estimated positions of the stationary points in the $(\dd \theta/\dd z^T, z)$ plane.
We suggest that this mismatch results from a reduced vertical shear stress in the stratified case considered here, in comparison with the uniform density fluid considered by the classical theory for a log-layer; buoyancy forces ``isolate'' the stacked shear layers from each other.

No attempt has been made to fit the model to the observations and, in order to obtain a better quantitative agreement, the analysis of section \ref{sec:theory} should be refined in the future.
In particular, a more realistic functional dependence of the mixing length on the HAB is needed, something which could be obtained only through a dedicated investigation, most likely in a controlled environment.
The geometric average is used here only for simplicity, due to the lack of a more physically motivated functional form.

A more realistic representation of the mixing length may also improve the dependence of the magnitude of the flux on the HAB.
If we compare figure \ref{fig:model} with figure \ref{fig:flux_scatter}-c, we see that the model predicts a marked decrease in the value of the maximum of the flux for higher HAB, while the observations indicate that this decrease is marginal at best.

Furthermore, the transport of turbulent kinetic energy should not be neglected, for which the full differential equations system need to be considered.
In this respect, it is interesting to note that, in the steady state, \ref{eq:B98}-a implies that the buoyancy flux is linear with depth.
Observations indicate this is not a realistic result (figure \ref{fig:flux_scatter}-d), suggesting that a dissipation or forcing term in the buoyancy gradient equation \ref{eq:B98}-a may be missing.
The derivation of the model from the Reynolds decomposed primitive equations would obviously provide a firmer ground for discussion.

Further progress also implies considering, in the forcing term $\Pc$ of equation \ref{eq:B98}, the effect of direct wave breaking, e.g.~due to wave energy concentration after reflection of a wave train.
This mechanism, whose energetic implications are discussed in some detail by \cite{scotti_biases_2015}, is not included in our forcing term, which rather assumes a classical wall bounded shear layer.
The forcing from breaking waves is likely related to the slope and horizontal structure of the sea-floor as well, another element which is ignored in our model, and which may contribute to the discrepancies with the observations noted above.

As a last remark, we briefly consider the small bump visible in the flux-gradient relation at $\dd\theta/\dd z^T\approx 5\times 10^{-2}\dC\m^{-1}$, in particular in panels b and c (filled symbols alone) of figure \ref{fig:flux_scatter}.}
This feature, not predicted by the model, may be linked to non-stationarities or, possibly, to a second instability of the flux-gradient relation, whose origin is at the moment unclear.
A preliminary investigation suggests that it may be more prominent in the profiles not having a layered structure.

Further progress in the analysis of this data-set implies a careful characterisation of the scales of the layers in the observations and, possibly, their initial development and further evolution.
While these issues go beyond the scope of this work, we expect that their correct description by the model will prove to be essential for its further development.
A closer description of the observations by the model may also provide the basis for applications, e.g.~as a turbulence closure for a larger scale ocean model.\\

We would like to thank the crew of the RV~Pelagia for enabling the deployment and recovery of the instruments, and all the technicians, in particular Martin Laan, indispensable for the success of the measurements. We would like to thank prof.~Leo Maas for providing comments and suggestions for the improvement of the model. The authors are grateful to the referees and the editor for providing insightful, detailed comments on the first version of this manuscript.

\appendix
\section{Appendix}
\label{sec:appendixA}

The velocity data recorded by the ADCP is used for transforming the measured temperature time series at each depth into a spatial series, assuming the validity of the ``frozen turbulence'' (Taylor's) hypothesis.
Since velocity is not constant, the approach of \citet{pinton_correction_1994} is followed.

The velocity field $v$ is decomposed into a mean component $\mean{v}$ and a fluctuating one $v'$ using a low-pass Butterworth filter with stop band $1/\sigma=3000\s$.
On average, this can be considered the forcing frequency, based on the frequency spectra of temperature and velocity fluctuations (not shown).
The mean velocity is higher in the upslope tidal phase, while the velocity fluctuations have a similar distribution during both tidal phases.

The mean velocity obtained by filtering the ADCP data is vertically averaged in four bins, and these averages are used to transform the temperature time series of each thermistor to the spatial domain (using the average velocity from the appropriate vertical bin).
The spatial series thus obtained has a non-constant sampling rate because velocity is not constant in time.
By linear interpolation to a constant time step of $0.2\m$, a constant step spatial series is obtained, reducing the size of the data set by a factor of approximately 2.
The vertical resolution is fixed to $0.7\m$ by the thermistor spacing.

Turbulence intensity ($\mean{v'^2}^{1/2}/\mean{v}$) estimated from the ADCP data is much smaller than $1$ most of the time, supporting the use of Taylors's frozen turbulence hypothesis.
However, since the ADCP resolution is insufficient to resolve the complete turbulence inertial range, this value of turbulence intensity is likely underestimated.
On the other hand, the presence of noise in the ADCP data artificially increases the estimated turbulence intensity.

Only the first increments of the Taylor transformed data are used in this work, for computing the horizontal component of the temperature gradient discussed in section \ref{sec:winters-d'asaro}.

\clearpage

\begin{table}
  \centering
  \begin{tabular}{l | l}
    $\nu$ & $10^{-6}\m^2\s^{-1}$ \\
    $u_*$ & $10^{-3}\m\s^{-1}$ \\
    $N_0$ & $10^{-3}\s^{-1}$ \\
    $\gamma$ & $10^{-3}$ \\
    $\delta$ & $10^{-1}$ \\
    $r$ & $50$
  \end{tabular}
  \caption[Model configuration]{Values of the dimensional and non-dimensional parameters used in the numerical solution of the model.
    \label{tab:model}}
\end{table}

\begin{table}
  \centering
  \begin{tabular}{l | l}
    Latitude & $36^\circ\, 58.885'\Nl$\\
    Longitude & $13^\circ\, 45.523'\Wl$\\
    Deepest thermistor depth& $2205\m$\\
    Deepest thermistor HAB& $5\m$\\
    Bottom slope & $9.4^\circ$ \\
    Number of thermistors & 144\\
    Thermistor vertical spacing & $0.7\m$\\
    Total length of array & $100.1\m$\\
    Total length of cable & $205\m$\\
    ADCP time resolution & $600\s$\\
    ADCP vertical resolution & $5\m$\\
    Deployment & 13 Apr 2013\\
    Recovery & 12 Aug 2013\\
  \end{tabular}
  \caption[Mooring description]{Description of the mooring from which the data is obtained.
    HAB is the height above the bottom.
    \label{tab:mooring}}
\end{table}

\clearpage

\begin{thebibliography}{32}
\expandafter\ifx\csname natexlab\endcsname\relax\def\natexlab#1{#1}\fi
\def\au#1{#1} \def\ed#1{#1} \def\yr#1{#1}\def\at#1{#1}\def\jt#1{\textit{#1}}
  \def\bt#1{#1}\def\bvol#1{\textbf{#1}} \def\vol#1{#1} \def\pg#1{#1}
  \def\publ#1{#1}\def\arxiv#1{#1}\def\org#1{#1}\def\st#1{\textit{#1}}

\bibitem[Baker \& Gibson(1987)]{baker_sampling_1987}
{\sc \au{Baker, M.~A.} \& \au{Gibson, C.~H.}} \yr{1987}  \at{Sampling
  {Turbulence} in the {Stratified} {Ocean}: {Statistical} {Consequences} of
  {Strong} {Intermittency}}.  \jt{J.~Phys.~Oceanogr.}  \bvol{17},
  \pg{1817--1836}.

\bibitem[Balmforth {\em et~al.\/}(1998)Balmforth, Llewellyn~Smith \&
  Young]{balmforth_dynamics_1998}
{\sc \au{Balmforth, N.~J.}, \au{Llewellyn~Smith, S.~G.} \& \au{Young, W.~R.}}
  \yr{1998}  \at{Dynamics of interfaces and layers in a stratified turbulent
  fluid}.  \jt{J.~Fluid Mech.}  \bvol{355},  \pg{329--358}.

\bibitem[Billant \& Chomaz(2000)]{billant_experimental_2000}
{\sc \au{Billant, P.} \& \au{Chomaz, J.-M.}} \yr{2000}  \at{Experimental
  evidence for a new instability of a vertical columnar vortex pair in a
  strongly stratified fluid}.  \jt{J.~Fluid Mech.}  \bvol{418},  \pg{167--188}.

\bibitem[Chalamalla \& Sarkar(2015)]{chalamalla_mixing_2015}
{\sc \au{Chalamalla, V.~K.} \& \au{Sarkar, S.}} \yr{2015}  \at{Mixing,
  {Dissipation} {Rate}, and {Their} {Overturn}-{Based} {Estimates} in a
  {Near}-{Bottom} {Turbulent} {Flow} {Driven} by {Internal} {Tides}}.
  \jt{J.~Phys.~Oceanogr.}  \bvol{45},  \pg{1969--1987}.

\bibitem[Cimatoribus \& {van Haren}(2015)]{cimatoribus_temperature_2015}
{\sc \au{Cimatoribus, A.~A.} \& \au{{van Haren}, H.}} \yr{2015}
  \at{Temperature statistics above a deep-ocean sloping boundary}.
  \jt{J.~Fluid Mech.}  \bvol{775},  \pg{415--435}.

\bibitem[Cimatoribus {\em et~al.\/}(2014)Cimatoribus, {van Haren} \&
  Gostiaux]{cimatoribus_comparison_2014}
{\sc \au{Cimatoribus, A.~A.}, \au{{van Haren}, H.} \& \au{Gostiaux, L.}}
  \yr{2014}  \at{Comparison of {E}llison and {T}horpe scales from {E}ulerian
  ocean temperature observations}.  \jt{J.~Geophys.~Res.-Oceans}  \bvol{119},
  \pg{7047--7065}.

\bibitem[Dillon(1982)]{dillon_vertical_1982}
{\sc \au{Dillon, T.~M.}} \yr{1982}  \at{Vertical overturns: {A} comparison of
  {Thorpe} and {Ozmidov} length scales}.  \jt{J.~Geophys.~Res.-Oceans}
  \bvol{87},  \pg{9601--9613}.

\bibitem[Gargett(1989)]{gargett_ocean_1989}
{\sc \au{Gargett, A.~E.}} \yr{1989}  \at{Ocean {Turbulence}}.
  \jt{Ann.~Rev.~Fluid Mech.}  \bvol{21},  \pg{419--451}.

\bibitem[Gregg(1987)]{gregg_diapycnal_1987}
{\sc \au{Gregg, M.~C.}} \yr{1987}  \at{Diapycnal mixing in the thermocline: {A}
  review}.  \jt{J.~Geophys.~Res.-Oceans}  \bvol{92},  \pg{5249--5286}.

\bibitem[Holford \& Linden(1999)]{holford_turbulent_1999}
{\sc \au{Holford, J.~M.} \& \au{Linden, P.~F.}} \yr{1999}  \at{Turbulent mixing
  in a stratified fluid}.  \jt{Dynam.~Atmos.~Oceans}  \bvol{30},
  \pg{173--198}.

\bibitem[Itsweire(1984)]{itsweire_measurements_1984}
{\sc \au{Itsweire, E.~C.}} \yr{1984}  \at{Measurements of vertical overturns in
  a stably stratified turbulent flow}.  \jt{Phys.~Fluids}  \bvol{27},
  \pg{764--766}.

\bibitem[Landau \& Lifshitz(1987)]{landau_fluid_1987}
{\sc \au{Landau, L.~D.} \& \au{Lifshitz, E.~M.}} \yr{1987} {\em Fluid
  {Mechanics}\/}, ,  \vol{vol.~6}.  \publ{Amsterdam (Netherlands): Elsevier}.

\bibitem[Martin \& Rehmann(2006)]{martin_layering_2006}
{\sc \au{Martin, J.~E.} \& \au{Rehmann, C.~R.}} \yr{2006}  \at{Layering in a
  {Flow} with {Diffusively} {Stable} {Temperature} and {Salinity}
  {Stratification}}.  \jt{J.~Phys.~Oceanogr.}  \bvol{36},  \pg{1457--1470}.

\bibitem[Mater \& Venayagamoorthy(2014)]{mater_quest_2014}
{\sc \au{Mater, B.~D.} \& \au{Venayagamoorthy, S.~K.}} \yr{2014}  \at{The quest
  for an unambiguous parameterization of mixing efficiency in stably stratified
  geophysical flows}.  \jt{Geophys.~Res.~Lett.}  \bvol{41},  \pg{4646--4653}.

\bibitem[Mater {\em et~al.\/}(2015)Mater, Venayagamoorthy, {St.~Laurent} \&
  Moum]{mater_biases_2015}
{\sc \au{Mater, B.~D.}, \au{Venayagamoorthy, S.~K.}, \au{{St.~Laurent}, L.} \&
  \au{Moum, J.~N.}} \yr{2015}  \at{Biases in {Thorpe} scale estimates of
  turbulence dissipation {Part} {I}: {Assessments} from large-scale overturns
  in oceanographic data}.  \jt{J.~Phys.~Oceanogr.}  \bvol{in press}.

\bibitem[Munk(1966)]{munk_abyssal_1966}
{\sc \au{Munk, W.~H.}} \yr{1966}  \at{Abyssal recipes}.  \jt{Deep-Sea Res.}
  \bvol{13},  \pg{707--730}.

\bibitem[Osborn \& Cox(1972)]{osborn_oceanic_1972}
{\sc \au{Osborn, T.~R.} \& \au{Cox, C.~S.}} \yr{1972}  \at{Oceanic fine
  structure}.  \jt{Geophys.~Fluid Dyn.}  \bvol{3},  \pg{321--345}.

\bibitem[Park {\em et~al.\/}(1994)Park, Whitehead \&
  Gnanadeskian]{park_turbulent_1994}
{\sc \au{Park, Y.-G.}, \au{Whitehead, J.~A.} \& \au{Gnanadeskian, A.}}
  \yr{1994}  \at{Turbulent mixing in stratified fluids: layer formation and
  energetics}.  \jt{J.~Fluid Mech.}  \bvol{279},  \pg{279--311}.

\bibitem[Phillips(1972)]{phillips_turbulence_1972}
{\sc \au{Phillips, O.~M.}} \yr{1972}  \at{Turbulence in a strongly stratified
  fluid---is it unstable?}  \jt{Deep-Sea Res.}  \bvol{19},  \pg{79--81}.

\bibitem[Pinton \& Labb\'e(1994)]{pinton_correction_1994}
{\sc \au{Pinton, J.-F.} \& \au{Labb\'e, R.}} \yr{1994}  \at{Correction to the
  {T}aylor hypothesis in swirling flows}.  \jt{J.~Physique {II}}  \bvol{4},
  \pg{1461--1468}.

\bibitem[Posmentier(1977)]{posmentier_generation_1977}
{\sc \au{Posmentier, E.~S.}} \yr{1977}  \at{The {Generation} of {Salinity}
  {Finestructure} by {Vertical} {Diffusion}}.  \jt{J.~Phys.~Oceanogr.}
  \bvol{7},  \pg{298--300}.

\bibitem[Prandtl(1935)]{prandtl_mechanics_1935}
{\sc \au{Prandtl, L}} \yr{1935}  \at{The {Mechanics} of {Viscous} {Fluids}}.
  \bt{In {\em Aeordynamics {Theory}\/} (ed. \ed{W.~F. Durand})}, ,
  \vol{vol.~3},  \pg{pp. 34--208}.  \publ{Berlin, Germany: Springer}.

\bibitem[Ruddick {\em et~al.\/}(2000)Ruddick, Anis \&
  Thompson]{ruddick_maximum_2000}
{\sc \au{Ruddick, B.}, \au{Anis, A.} \& \au{Thompson, K.}} \yr{2000}
  \at{Maximum likelihood spectral fitting: {The} {Batchelor} spectrum}.
  \jt{J.~Atmos.~Ocean.~Tech.}  \bvol{17},  \pg{1541--1555}.

\bibitem[Ruddick {\em et~al.\/}(1989)Ruddick, McDougall \&
  Turner]{ruddick_formation_1989}
{\sc \au{Ruddick, B.~R.}, \au{McDougall, T.~J.} \& \au{Turner, J.~S.}}
  \yr{1989}  \at{The formation of layers in a uniformly stirred density
  gradient}.  \jt{Deep-Sea Res. Pt A}  \bvol{36},  \pg{597--609}.

\bibitem[Scotti(2015)]{scotti_biases_2015}
{\sc \au{Scotti, A.}} \yr{2015}  \at{Biases in {Thorpe}-{Scale} {Estimates} of
  {Turbulence} {Dissipation}. {Part} {II}: {Energetics} {Arguments} and
  {Turbulence} {Simulations}}.  \jt{J.~Phys.~Oceanogr.}  \bvol{45},
  \pg{2522--2543}.

\bibitem[Thorpe(1977)]{thorpe_turbulence_1977}
{\sc \au{Thorpe, S.~A.}} \yr{1977}  \at{Turbulence and {Mixing} in a {Scottish}
  {Loch}}.  \jt{Royal Society of London Philosophical Transactions A}
  \bvol{286},  \pg{125--181}.

\bibitem[{van Haren} {\em et~al.\/}(2015){van Haren}, Cimatoribus \&
  Gostiaux]{van_haren_where_2015}
{\sc \au{{van Haren}, H.}, \au{Cimatoribus, A.} \& \au{Gostiaux, L.}} \yr{2015}
   \at{Where large deep-ocean waves break: {Where} large deep-ocean waves
  break}.  \jt{Geophys.~Res.~Lett.}  \bvol{42}~(7),  \pg{2351--2357}.

\bibitem[{van Haren} {\em et~al.\/}(2009){van Haren}, Laan, Buijsman, Gostiaux,
  Smit \& Keijzer]{van_haren_nioz3:_2009}
{\sc \au{{van Haren}, H.}, \au{Laan, M.}, \au{Buijsman, D.-J.}, \au{Gostiaux,
  L.}, \au{Smit, M.~G.} \& \au{Keijzer, E.}} \yr{2009}  \at{{NIOZ3:}
  independent temperature sensors sampling yearlong data at a rate of 1 {Hz}}.
  \jt{{IEEE} J.~Oceanic Eng.}  \bvol{34},  \pg{315--322}.

\bibitem[Warhaft(2000)]{warhaft_passive_2000}
{\sc \au{Warhaft, Z.}} \yr{2000}  \at{Passive scalars in turbulent flows}.
  \jt{Ann.~Rev.~Fluid Mech.}  \bvol{32},  \pg{203--240}.

\bibitem[Winters \& D'Asaro(1996)]{winters_diascalar_1996}
{\sc \au{Winters, K.~B.} \& \au{D'Asaro, E.~A.}} \yr{1996}  \at{Diascalar flux
  and the rate of fluid mixing}.  \jt{J.~Fluid Mech.}  \bvol{317},
  \pg{179--193}.

\bibitem[Wunsch \& Kerstein(2001)]{wunsch_model_2001}
{\sc \au{Wunsch, S.} \& \au{Kerstein, A.}} \yr{2001}  \at{A model for layer
  formation in stably stratified turbulence}.  \jt{Phys.~Fluids}  \bvol{13},
  \pg{702--712}.

\bibitem[Zhou \& Gao(1997)]{zhou_confidence_1997}
{\sc \au{Zhou, X.~H.} \& \au{Gao, SUJUAN}} \yr{1997}  \at{Confidence intervals
  for the log-normal mean}.  \jt{Stat.~Med.}  \bvol{16},  \pg{783--790}.

\end{thebibliography}
\end{document}